# Thermal Effects in Stretching of Go-like Models of Titin and Secondary Structures

**Marek Cieplak,[1,2] Trinh Xuan Hoang,[3,4] and Mark O. Robbins[2]**

[1] *Institute of Physics, Polish Academy of Sciences, Al. Lotników 32/46, 02-668 Warsaw, Poland*
[2] *Department of Physics and Astronomy, The Johns Hopkins University, Baltimore, MD 21218*
[3] *The Abdus Salam International Center for Theoretical Physics, Strada Costiera 11, 34100 Trieste, Italy*
[4] *Institute of Physics, NCST, 46 Nguyen Van Ngoc, Hanoi, Vietnam*

Correspondence to:
Marek Cieplak,
Institute of Physics,
Polish Academy of Sciences,
Al. Lotników 32-46
02-668 Warsaw, Poland
Tel: 48-22-843-7001
Fax: 48-22-843-0926
E-mail: mc@ifpan.edu.pl

Grant sponsors: NSF DMR-0083286, and KBN (Poland) – 2 P03B-032-25.

# Abstract

The effect of temperature on mechanical unfolding of proteins is studied using a Go-like model with a realistic contact map and Lennard-Jones contact interactions. The behavior of the I27 domain of titin and its serial repeats is contrasted to that of simple secondary structures. In all cases thermal fluctuations accelerate the unraveling process, decreasing the unfolding force nearly linearly at low temperatures. However differences in bonding geometry lead to different sensitivity to temperature and different changes in the unfolding pattern. Due to its special native state geometry titin is much more thermally and elastically stable than the secondary structures. At low temperatures serial repeats of titin show a parallel unfolding of all domains to an intermediate state, followed by serial unfolding of the domains. At high temperatures all domains unfold simultaneously and the unfolding distance decreases monotonically with the contact order, that is the sequence distance between the amino acids that form the native contact.

## INTRODUCTION

The giant molecule titin is one of the prime objects of mechanical studies of single biological molecules, because of its role in controlling the degree of extension and elasticity of smooth, skeletal, and cardiac muscles.[1-5] Titin is known to consist of many (~300) globular domains which are connected in series. The domains are of similar structure but different homology. The first domain to have its native conformation determined was the 27$^{th}$ immunoglobulin domain of the I band of titin, I27, which has the structure of a β-sandwich.[6] The structures of a growing number of other domains have been determined,[7-10] many of which contain short α-helix regions in addition to β-sheets. Stretching studies of both natural and engineered titin have been accomplished by a variety of techniques.[11-20] All reveal sawtooth patterns in the force ($F$) – tip displacement ($d$) curves that are consistent with a predominantly serial unraveling of domains.[13,21] The reason is that the bonds that require the largest force to break rupture near the start of the unraveling process of each domain.

All atom simulations of a single I27 domain in water have reproduced many experimental properties and helped to interpret them. For example, the bonds that require the largest force to break have been identified as six hydrogen bonds,[22,23] and the structure of an intermediate state that forms during unfolding has been identified.[24] However, these studies are limited to very high velocities and can not easily address the behavior of multiple domains. In addition, their computational expense makes it difficult to explore generic features of unfolding in a wide range of proteins, in order to develop a more global theoretical understanding.

Coarse-grained Go-like models[25,26] that only use structural information about proteins are able to capture many of their properties with minimal computational effort. In recent papers we have contrasted the folding and mechanical unfolding behavior of typical secondary structures[27] and then of a titin domain and its tandem repeats[21] as modeled by a Go-like system.[25,26] The focus of these studies was on establishing scenarios of unfolding, as determined by the order of bond breaking, and investigating their relationship to contact formation during thermal folding. In general, there is no correlation, because of the crucial role the geometry of loading plays in unfolding.

Our previous mechanical unfolding studies were effectively done at zero temperature in order to minimize fluctuations and rate dependence. The purpose of this paper is to investigate the effects of thermal fluctuations on unfolding of proteins. While temperature can not be varied dramatically in an aqueous environment, the effective strength of thermal fluctuations can be varied experimentally through changes in solute concentrations. This does not yet appear to have been done systematically, but we find it can produce marked changes in unfolding scenarios. We begin by considering two simple secondary structures: $\alpha$-helices and $\beta$-sheets. Then single domains and multiple repeats of titin are studied. Thermal fluctuations aid unfolding in all cases, decreasing both the unfolding force and the extension at unfolding. At low temperatures, the rate of these changes and their effect on the unfolding pattern, depend strongly on the geometry of the protein. In some cases, the stiffness of the mechanical device can also be important. However, at high temperatures the unfolding scenarios become universal and the force can be described by an entropic worm-like-chain model.[28,29]

The remainder of the paper is organized as follows. The following section discusses the Go-like model that we use and the details of our molecular dynamics simulations. The next section examines the behavior of secondary structures and titin. Changes induced by coupling many domains and varying velocity are discussed for titin. The final section presents a summary and conclusions.

## MODEL AND METHOD

The models we use are coarse-grained and Go-like,[25,26] and were initially developed for studies of folding. Each amino acid is represented by a point particle of mass *m* located at the position of the $C_\alpha$ atom. The interaction potential is constructed so that the native structure minimizes the potential energy. The original version is described in Refs. 30-32 and was used in our previous studies of stretching.[21,27] Here we use the improved model described in Ref. 33, and emphasize the improvements in the following discussion.

The interactions between amino acids in the native structure of a protein are divided into native and non-native contacts. Instead of adopting a uniform cutoff criterion of 7.5Å between $C_\alpha$'s in a native contact,[21,27] we follow the procedure of Tsai et al.[34] and base the criterion on overlap of atoms in the amino acids. Atoms are represented by spheres whose radii are a factor of 1.24 larger than the atomic van der Waals radii to account for the softness of the potential. With this definition of contact, the separation between $C_\alpha$ atoms in native contacts varies from 4.3 to 12.8 Å. Using this more accurate potential has strong effects on the folding kinetics,[33,35] but we show below that the stretching curves for titin and secondary structures are relatively unaffected.

In our studies of secondary structures we consider "synthetic" geometries of two generic structures; an $\alpha$-helix and $\beta$-sheet.[27] The native structure of I27 is taken from the PDB[36] data bank where it is stored under the name 1tit which we shall use as an alternative to I27. 1tit consists of 89 residues which are organized into eight β-strands and connecting turns (Fig. 8). As in Ref. 27, tandem structures of two or more domains are constructed by repeating 1tit domains in series with one extra peptide link between the domains. Experimental links may be one or two amino acids longer and modify the structure of the terminal groups.[15,19] However, no structural information is available from which to build a Go-like potential for the links and our results are insensitive to small changes in length.

The interactions between amino acids are modeled by a 6-12 Lennard-Jones potential $4\varepsilon[(\sigma_{ij}/r_{rj})^{12}-(\sigma_{ij}/r_{rj})^{6}]$, where $r_{ij}$ is the distance between $C_\alpha$ atoms $i$ and $j$. All contacts have the same energy scale ε. The characteristic length $\sigma_{ij}$ is adjusted so that the energy minimum for each native contact coincides with the distance between $C_\alpha$ atoms in the native structure. For all pairs that do not form a native contact, $\sigma_{ij} = \sigma = 5$Å, and the potential is truncated at $2^{1/6}\sigma$ to produce a purely repulsive interaction. As shown below, the 10-12 Lennard-Jones potential gives very similar trends.

Neighboring $C_\alpha$ atoms are tethered by a strong potential with a minimum at the peptide bond length of 3.8Å. In the following we use a harmonic bond with force constant $100\varepsilon/\text{Å}^2$, because the strongly anharmonic bond used in Refs. 21 and 27 stretched too easily at small forces. This shifted the positions of force peaks relative to all-atom calculations.[21,22] A four-body term that favors the native sense of chirality is also introduced.[33] It vanishes for the native chirality and has an energy penalty of ε for the opposite chirality.

The effective temperature is given by the ratio $\tilde{T} = k_B T/\varepsilon$, where $T$ is the temperature and $k_B$ is the Boltzmann's constant. The energy scale ε represents the typical binding energy of native contacts and includes hydrogen bonding, hydrophobic, and van der Waals interactions. In principle, different native contacts have different binding energies, but since our model includes only structural information, there is no simple mechanism for introducing variations in energy. Estimates for ε in real proteins[37] range from 0.07 to 0.2eV, giving $\tilde{T}$ = 0.1 to 0.3 at room temperature. The value of $\tilde{T}$ can be varied slightly by changing $T$, and over a wide range by changing solvent conditions to vary ε. We will express all our results in terms of $\tilde{T}$, keeping in mind that when we talk about increasing this effective temperature the physical temperature $T$ may remain constant.

A Langevin thermostat[38] with damping constant γ is coupled to each $C_\alpha$ to control $\tilde{T}$. For the results presented below $\gamma = 2m/\tau$, where $\tau = \sqrt{m\sigma^2/\varepsilon}$ ~ 3ps is the characteristic time for the Lennard-Jones potential. This is large enough to produce the overdamped dynamics appropriate for proteins in a solvent,[33] but roughly 25 times smaller than the realistic damping from water.[39] Previous studies show that this speeds the diffusive dynamics without altering behavior, and tests with larger γ confirm a linear scaling of folding times with γ.[30,31] Thus the folding times for titin reported below should be multiplied by 25 for comparison to experiment. The effect of γ on mechanical unfolding is discussed below (Fig. 17).

In our simulations of stretching, both ends of the protein are attached to harmonic springs of spring constant $k$. Since the two springs are in series, this is equivalent to using an atomic force microscope (AFM) cantilever with stiffness $k/2$. As in Refs.21 and 27 we refer to simulations with $k = 30\varepsilon/\text{Å}^2$ and $k = 0.12\varepsilon/\text{Å}^2$ as stiff and soft, respectively. Generally, the stiffer the cantilever, the more intricate the sequence of unfolding events.[21,27] Using $\varepsilon = 0.2eV$, the soft and stiff cases correspond to cantilevers of stiffness 0.2 to 48 N/m. Typical AFM cantilevers are closer to the soft case, ranging from 0.06 to 0.6N/m. Note however, that if $\tilde{T}$ is increased by lowering ε, our stiff cantilever becomes closer to experimental stiffnesses. The effective stiffness of the system is reduced by the compliance of the protein itself. This is particularly important in experiments where a large number of domains are stretched in series.[15,20]

Stretching is implemented parallel to the initial end-to-end vector of the protein. The outer end of one spring is held fixed at the origin, and the outer end of the other is pulled at constant speed $v_p$. The separation of the moving end from the origin would correspond to the cantilever displacement in an AFM. The separation in the native state is used as the zero for the displacement $d$. Unless otherwise noted $v_p$= 0.005 Å/τ, since little velocity dependence was observed below this $v_p$ in our low temperature studies.[21] This corresponds to a velocity of about $7\times10^6$ nm/s when the small value of γ is taken into account. This is orders of magnitude faster than velocities in AFM experiments[13,18,40] (1 to $10^4$nm/s), but much slower than all atom simulations where $v_p$ is $10^{10}$nm/s or greater.[23]

The folding and unfolding processes are characterized by the order in which native contacts are formed or broken. At finite temperatures, contacts may break or form many times due to thermal fluctuations. When discussing folding, we determine the average time $t_c$ for each contact to form for the first time. When discussing the succession of rupturing events, we determine average cantilever distance, $d_u$, at which a contact exists for the last time. As a technical criterion for the presence of a contact between amino acids $i$ and $j$ we take the $C_\alpha$–$C_\alpha$ distance not to exceed $1.5\sigma_{ij}$. The folding data were averaged over 500 trajectories and the stretching data over 20 trajectories because fluctuations are smaller. Throughout the paper, the symbol sizes are measures of the statistical error bars. The equations of motion are integrated using a fifth-order predictor-corrector algorithm with time step $dt$=0.005τ.

Unfolding is also characterized by the force - displacement curves that would be measured by, for example, an AFM. The force $F$ is determined from the extension of the pulling spring, and the displacement $d$ from the change in the position of its outer end relative to the native state. At finite temperature there are substantial fluctuations in $F$ due to excitations of the pulling spring and protein. The force is averaged over 100τ to reduce this random noise. For $v_p$= 0.005Å/τ this corresponds to an averaging distance $d_A$ of 0.5Å. Figure 1 illustrates that $d_A$=0.5Å is small enough to retain the structure due to breaking of native contacts at $\tilde{T} = 0$, and large enough to reduce the rapid thermal fluctuations at $\tilde{T} = 0.2$. Here a single domain of titin is stretched by the stiff spring and only the region near the first force peak is shown.

To conclude the methods section, Figure 2 compares the *F–d* curve at $T = 0$ for the potential model used in the remainder of the paper to two other models. The top panel compares to the results from our previous studies.[21,27] As mentioned above, the anharmonic tethering potential leads to excessive stretching of the protein that shifts force peaks to larger *d*. The fine structure is also changed because native contacts were identified using a uniform cutoff criterion. However, the main force peaks have similar magnitudes and involve similar regions of the protein. The bottom panel shows that changing the contact potential from a 6-12 to 10-12 Lennard-Jones potential has surprisingly little effect on the force curve. Some have argued that this potential gives a more accurate description of hydrogen bonds.[41] Increasing temperature reduces the differences between potentials and eventually the behavior becomes universal.[42]

## RESULTS AND DISCUSSION

### Stretching of secondary structures

No experiments on stretching of secondary structures have yet been reported but it is instructive to consider their behavior theoretically. Here we focus on the basic building blocks of most proteins: α-helices and β-hairpins. As shown previously,[27] their different geometries lead to very different distributions of stress and resulting unfolding behavior.

Figures 3 and 4 show the force-displacement curves for the α-helix with stiff and soft springs, respectively. The $\tilde{T} = 0$ behavior is typical of all proteins.[27] *F* shows a series of upward ramps where the cantilever stretches and the protein retains its configuration. Each ramp terminates in a peak when some contacts break. This produces a rapid drop in force as the protein unravels and the extension of the cantilever is reduced. The stiffer the spring, the greater the drop in force during each rupture, and the more rugged the *F–d* curve. The softer the spring, the more contacts may break in a single event.

When the α-helix is stretched along its length, the stress is carried in series by all native contacts that connect adjacent turns of the helix. This means that at $\tilde{T} = 0$ bonds will break in order of their strength. The dominant contact associated with each peak corresponds to the hydrogen bond between turns that would connect beads $i$ and $i+4$, but contacts to $i+2$ and $i+3$ break at the same time. Failure begins at the ends, where these weaker contacts are absent, and propagates inward.[27] Each peak is doubled because there are identical groups of contacts on the two ends of the helix. The central bonds are the hardest to break and give the final, highest peaks.

At finite $\tilde{T}$, thermal fluctuations assist contact rupture. Contacts rupture at smaller forces, and correspondingly smaller distances. Figure 5 shows the decrease in the maximum peak height $F_{max}$ and its position $d_{max}$ as $\tilde{T}$ increases. Both quantities change roughly linearly at low $\tilde{T}$. For $\tilde{T} > 0.15$ there is no longer a clear peak in *F*, just a broad plateau whose height is plotted in Fig. 5. The plateau reflects a major change in the unfolding process. For $\tilde{T} = 0.1$, the fluctuations between different runs are small (Fig. 3), and the order of bond breaking is close to that at $\widetilde{T} = 0$. At $\widetilde{T} = 0.2$ (Figs. 3 and 4), bonds do not break in a fixed order. Indeed, each contact breaks and reforms several times along the plateau.

The plateau in $F - d$ at $\widetilde{T} = 0.2$ can be understood from a very simple model. For the α-helix, the pulling force is carried by all contacts and increases the probability that each will break. Moreover, any thermally broken contact allows the whole protein to extend. The situation is very similar to a linear string of bonds that have two metastable states of different length. The shorter corresponds to the native contact and has a lower free energy. The longer corresponds to the broken contact. In the absence of an applied force, the probability of a broken contact scales as the exponential of the free energy difference between the two states. An applied force simultaneously lowers the free energy difference for all bonds. Initially, the probability of a single broken bond rises and the average length increases slightly. At the force where the free energy difference vanishes, the two states can coexist in any proportion. The chain of bonds will expand at constant force by increasing the fraction of broken bonds. The situation is analogous to liquid/gas coexistence where the volume expands at fixed pressure. As expected from this analogy, the force plateau decreases in width as $\widetilde{T}$ increases, and disappears above $\widetilde{T} = 0.4$.

The temperatures where changes in unfolding behavior are observed correlate with characteristic temperatures obtained from previous equilibrium studies of folding.[33] The probability for the α-helix to be in the completely folded state drops with increasing temperature and reaches one half at $\widetilde{T}_f = 0.24$. Thus it is not surprising that an applied force leads to fluctuations in unfolding patterns at the slightly lower temperature of $\widetilde{T} = 0.2$. The equilibrium structure of the protein at $\widetilde{T}_f$ remains close to the native state. A measure of the temperature where the protein unfolds to a random state is given by the location $\widetilde{T}_{\max}$ of the maximum in the specific heat.[43] For the α-helix, $\widetilde{T}_{\max} = 0.36$, and there is little evidence of a force plateau at higher temperatures in Figs. 3 and 4. Indeed the force curves for $\widetilde{T} = 0.6$ can be fit to the worm-like-chain (WLC) model,[28,29] which ignores contact energies and focuses on the entropy associated with different configurations of the protein. Ref. 42 examined several different proteins and found in each case that the force approached the entropic limit for $\widetilde{T}$ above $\widetilde{T}_{\max}$.

Figures 6 and 7 illustrate the behavior of a 16-acid β-hairpin B16. There is a 180$^{\circ}$ bend in the middle of the protein. The two sides are connected by a series of rung-like hydrogen bonds between opposing pairs of amino acids. When the ends are pulled apart, the force is localized on the unbroken contact closest to the ends. When it ruptures, stress is transferred to the next contact and so on. The $\widetilde{T} = 0$ force curve shows a periodic series of peaks and dips as each rung breaks in turn. Slight differences between rungs lead to an alternation in the spacing between peaks. As for the α-helix, increasing $\widetilde{T}$ lowers the force peaks (Fig. 6 inset) and shifts them to smaller *d*. However, the evolution of the unfolding pattern is very different.

Equilibrium studies show that B16 spends half of the time in the native state at $\widetilde{T}_f = 0.07$. This is much lower than the corresponding temperature for H16, but the fluctuations from the native state have less effect on the end-to-end distance. Only when the unbroken bond closest to the ends breaks does $d$ increase. Moreover, an applied stress only shifts the free energy of this terminal bond. Thus unfolding follows the same pattern as at $\widetilde{T} = 0$ until the temperature is close to $\widetilde{T}_{\max} = 0.9$. However, the bonds may break and reform many times, depending on the spring stiffness.

Simulations with the stiff spring essentially fix the end-to-end distance of the protein. For $\widetilde{T} << \widetilde{T}_{\max}$ this is mainly determined by the number of contacts closest to the ends that have been broken. $F$ is small when $d$ corresponds to a configuration where the remaining bonds are unstressed and $F$ is large when $d$ corresponds to a stressed configuration. These oscillations remain fairly strong for $\widetilde{T} = 0.2$. By $\widetilde{T} = 0.6$ bonds near the ends break and form multiple times and the oscillations are averaged out. Simulations with the soft spring are closer to a constant force[13] ensemble. There is a pronounced plateau in $F$ that corresponds to coexistence of broken and native states of the last bond(s). While this is similar to the coexistence seen for the α-helix, it only applies to the last bond rather than all the bonds along the length. Thus the pattern of unfolding is always the same. The serial nature of fluctuations in the α-helix makes its behavior less sensitive to cantilever stiffness.

**Folding of one domain of titin**

A ribbon representation of the domain I27 is shown in Figure 8. The main force peak seen in Figs. 1 and 2 corresponds to the rupture of contacts between the strands A' and G combined with A and G. These contacts are the longest ranged and might be expected to form last on folding but this is not the case in our simulations.[21]

Figure 9 shows the median folding time as a function of $\widetilde{T}$. The U-shaped curve is much narrower than the one found for the less accurate, uniform cutoff model considered in reference.[21] The temperature of optimal folding, $\widetilde{T}_{\min}$, is 0.275 $\varepsilon/k_B$. The median folding time at $\widetilde{T}_{\min}$ is 3800 τ and the temperature where the time has doubled $\widetilde{T}_{g2}$ is 0.22 $\varepsilon/k_B$. The folding temperature $\widetilde{T}_f \approx 0.2$ is lower than $\widetilde{T}_{\min}$. This suggests poorer folding properties[33,44] than in the uniform cutoff model[21] where $\widetilde{T}_f > \widetilde{T}_{\min}$. However, $\widetilde{T}_f$ and $\widetilde{T}_{\min}$ are quite comparable, and the definition of the native basin (all native contacts at distances less than 1.5 $\sigma_{ij}$) is not very precise. As noted above, unfolding is more closely associated with the maximum in the specific heat. The inset of Fig. 9 shows that the value of $\widetilde{T}_{\max} = 0.8$ is much larger than $\widetilde{T}_f$ and $\widetilde{T}_{\min}$. All these facts suggest that our model provides a reasonable description of folding.

The succession of folding events, as measured by average times $t_c$ to form contacts, is shown in Figure 10. Previous work suggests a one-to-one correspondence between $t_c$ and contact order defined as the sequence distance $|j-i|$.[31] While $t_c$ tends to rise monotonically with contact order, there are pronounced exceptions. In particular, the group of contacts between strands C and F with $|j-i|$ near 40 gets established at substantially longer times than all remaining contacts. This feature is independent of the version of the Go model, including the uniform contact model,[21] models with 10-12 contact potentials, and models with dihedral angle terms such as those of Clementi et al..[41] This indicates that deviations from the correlation between folding times and contact order may be robust.

To date, all-atom simulations are too computationally intensive to consider folding of titin. However, Paci and Karplus[45] considered thermal unfolding from the native state after a sudden increase in the temperature to 450K. While they only present results for a single simulation, it is interesting to note that the first bonds to break in their simulations are the same C+F bonds that form last in our model. The higher contact order A'+G and B+G bonds that break last in mechanical unfolding (see below) remain intact until much later in the thermal unfolding process. The most stable contacts seem to involve B, E and D, which form before the highest contact order bonds in our folding studies. It would be interesting to compare our folding results at higher temperatures to all-atom unfolding simulations with greater statistics.

Fowler and Clarke have used amino-acid substitutions and Φ-value analysis to study folding of 1tit.[46] They find very low Φ-values for strands A, A' and G, indicating that they form after the transition state. They identify the core of the transition state with bonds between acids in the center of strands B, C, E and F. This suggests that C+F bonds form early in the folding process. This difference from our results could reflect the simple nature of our Go-model, which can not be expected to capture all important interactions. However, Paci and Karplus[45] also found that C+F bonds are less stable. It is interesting to note that the high Φ-value for C at i=34 was not measured directly, but inferred from the geometry and values for nearby mutations.[46] The measured Φ-value for the D strand is actually somewhat higher, which would be more consistent with our folding sequence. We are exploring another possibility for the difference between experimental and theoretical results, which is that Φ-value analysis and average first contact times may give different sequences due to the wide distribution of contact times.

**Stretching of one domain of titin**

Force versus displacement curves for the I27 domain of titin are shown in Figures 11 and 12 for the soft and stiff pulling springs, respectively. As usual, the stiff curves are more structured, but both reveal two major peaks at low $\tilde{T}$. Fig. 13 shows the dependence of the peak heights and positions on $\tilde{T}$. Both peaks shift to lower forces and displacements with increasing $\tilde{T}$, and the additional structure observed with the stiff spring is gradually smeared out. The height of the first peak drops nearly linearly, and the initial change in the second peak height is linear. Both are nearly independent of stiffness and averaging interval. By $\tilde{T}_{max} = 0.8$ the entire force curve can be fit to the purely entropic WLC form. This is illustrated below for the more dramatic case of many domains (Fig. 16).

While the above changes with $\tilde{T}$ are similar to those found for secondary structures, the peaks for titin persist to much higher temperatures. Sharply defined peaks remain visible for $\tilde{T}$ close to $\tilde{T}_{max}$ in titin, while the peaks for secondary structures were smeared out for $\tilde{T}$ greater than 0.1. The heights of the force peaks are also substantially greater, even though the binding energy and form of the contact potential are the same. These differences between titin and the secondary structures are intimately related to its geometry and the resulting unfolding pattern.

We illustrate the unfolding scenarios by plotting the last distance at which the contact is found, $d_u$, against the contact order. Results for the soft and stiff springs are shown in the top and bottom panels of Figure 14, respectively. Bonds between each pair of the strands shown in Fig. 8 are indicated by different symbols that are identified in the $\tilde{T} = 0$ plots. For example, bonds between A and G, referred to as A+G, are indicated by closed circles.

When the stiff spring is used, bond breaking occurs in a steady series of separate events that is consistent with all atom simulations.[21,22] The highest contact order bonds coupling A, A' and G break first and are associated with the first force peak. Some bonds connecting A+B and B+G break at about the same time and contribute to the first peak. Others, and A'+B and F+G bonds, break between the two peaks. The C+F contacts fold last, but get ruptured in the middle of the process at the second force peak. B+E bonds break slightly later at $\tilde{T} = 0$ and make a small contribution to the second peak. The D+E bonds are last to unravel, and short-ranged contacts unravel throughout the process. The order is similar when the soft spring is used, but many more bonds break simultaneously. This is most pronounced at the first peak, which includes F+G, A'+B, and some B+E bonds.

As the temperature increases, the unfolding scenarios simplify. For most bonds $d_u$ decreases with increasing $\tilde{T}$, but $d_u$ increases for some bonds. At $\tilde{T} = 0.2$ the same groups of bonds remain associated with the two main peaks: A+G, A+G' and A+B bonds break at the first peak, and C+F and B+E at the second peak. However, the C+F bonds now break after the B+E bonds for both soft and stiff springs. Another change is found in the stiff spring results, where the F+G bonds break after the second peak instead of before. As $\tilde{T}$ increases further, the range of $d_u$ at a given contact order decreases. By $\tilde{T}_{max} = 0.8$ (Fig. 14) the results for both springs approach the entropic limit, where $d_u$ drops monotonically with increasing contact order. Similar curves are obtained for other proteins at $\tilde{T}_{max}$.[42] In this limit, the order of unraveling is nearly inverse to the folding order at $\tilde{T}_{min}$.

Previous work has focussed attention on the role of six hydrogen bonds connecting A' and G in stabilizing titin.[22,23,45,49] These bonds, and A+G bonds, represent the most direct path of stress transfer between the two ends of the domain, and break during the first peak in all models. However, the A+G bonds break before the peak, producing a small shoulder at forces of order 2ε/Å in both Figs. 11 and 12. Similar shoulders have been found in experiments[14] and identified with A+G bonds using atomistic simulations and genetic mutations.[22,24] The configuration with A+G bonds broken is called the intermediate (I) state, and in AFM experiments unfolding occurs from this state. Recent work shows that at the lower rates typical of physiological conditions unfolding occurs directly from the native state at a force lower than the shoulder.[40]

The bonds that are most important in producing the large force peak in Figs. 11 and 12 are the A'+G bonds. They run perpendicular to the applied force, while the key bonds in the secondary structures described above had a substantial component along the pulling force. As a result, all A'+G contacts can contribute in parallel to carrying the stress. In contrast, each bond was a potential failure site in the α-helix and only one bond carried the stress in the β-hairpin. These geometric factors are important in producing large and thermally stable force peaks for titin, but we find that other bonds are also important in determining the height of the first peak.

The contributions of different bonds to the force can be determined by eliminating the attractive forces between some native contacts and reevaluating the force curves. Figure 15 shows the region of the first force peak from stiff spring simulations at $\tilde{T} = 0$ and 0.2 that include different subsets of native contacts. Removing all bonds involving the A strand has relatively little effect on the height of the maximum at $d$ = 15Å. However, the shoulder from 6 to 10 Å disappears because it is associated with breaking A+G bonds in the native state. Removing A completely from the protein (not shown) has the same effect on the peak height as removing all bonds involving A, but shifts the curve to smaller $d$. These results are consistent with experiments and simulations[24] where the A strand was completely removed from the protein.

Figure 15 also shows force curves where only A'+G bonds are included. The peak force is reduced from the native result by a factor of two at both temperatures. Thus other bonds are important in stabilizing titin. We explored adding different sets of bonds to determine their role. Adding bonds between any one or two pairs of strands has little effect on the peak force. For example, the bonds from A or A' to B and B to G are the only sequences of bonds that connect A or A' directly to other strands and then to G. However, they produce a relatively small increase in peak force. This suggests that the A'+G bonds transmit most of the force at the main peak, and that other bonds contribute indirectly by stabilizing the geometry of the A' and G strands. To test this we considered a model where all bonds coupling A', B, F and G to each other were included, but all bonds involving C, D and E were not. As expected, the force peak rises most of the way to the result for the case where A is excluded.

The above conclusions are consistent with recent work that combined experimental and all-atom simulation studies[47] of mutant forms of 1tit. Mutations in the C, D and E regions had little effect on the unfolding force. Mutations that affected the B, F and G strands produced a larger effect, indicating that they are important in stabilizing the transition state. Of course mutations that affect A'+G contacts directly, have an even larger effect.

Note that after the first force peak, strands G, A and A' become taut and the protein rotates so the force is applied to the ends of B and F. These strands are not coupled directly, but through the rest of the protein. As at the first peak, the contacts are predominantly perpendicular to the pulling stress and act in parallel to produce a stronger bond.

**Stretching of several domains of titin**

We analysed stretching of several domains at $\tilde{T} = 0$ in Ref. 21. The domains were found to unwind in series, one by one. The resulting force curve (Fig. 16) is a sequence of nearly identical repeats of the force curve for a single domain. This serial unfolding is not observed for all multidomained proteins.[27] It arises in proteins like titin and T4 lysozyme 1b6i[42,48] where the largest force peak for a single domain occurs near the beginning of unfolding. The domains then remain largely intact as the force rises to each major peak. At the peak, just one domain breaks, releasing the stress on the others and allowing the process to repeat. Parallel unfolding is found for proteins like the α−helix where the force curve for a single domain rises gradually during unfolding.

In general, there is a complex mixture of parallel and serial unraveling events. Indeed, as noted above, the A+G bonds in titin break before the main peak. The $\tilde{T} = 0$ results in Fig. 16 show a shoulder before each peak that gets narrower as more domains unfold. At this shoulder, the A+G bonds on all folded domains become unstable and break in parallel. Thus all domains are in the I state as the force rises further to the maximum. The same shoulder was observed in experiments and used to infer the presence of the I state.[14] The increased length of the I state, 6.6 Å, was determined from the rise in the width of the shoulder with the total number of domains. We find an increased length of 5 Å, using the same procedure. However, this represents the change from the stressed native state to I. We find the change from the unstressed native state to I is about 11 Å.

Figure 16 shows how temperature changes the force curves for 5 repeats of titin (5tit). The soft spring is used, since it corresponds more closely to experimental cantilevers. Serial unfolding is observed at low temperatures. Results for $\tilde{T} = 0.2$ and 0.3 are close to periodic repeats of Figs. 11 and 12. However, the second peak is sometimes absent at $\tilde{T} =$ 0.2 and only seen during unfolding of the first domain at $\tilde{T} = 0.3$. These temperatures should be comparable to those in experiments, where the second peak is not observed. The lower rates of experiments may also be important in suppressing the second peak. They may also shift the changes in unfolding sequence of Fig. 14 to lower temperatures.

For $\tilde{T} \geq \tilde{T}_{\max} = 0.8$, the force rises monotonically with no evidence of serial structure. In this entropic limit, thermal fluctuations are strong enough to break all domains with no applied force. As a result, the repeated units behave like a single long chain and the entire curve can be fit to the WLC model. The dashed line through the $\tilde{T} = 0.8$ results shows a fit to the WLC force, $F = (k_B T / 4p)[(1 - d_{1,N}/L)^{-2} - 1 + 4d_{1,N}/L]$ as a function of the end-to-end length of the protein, $d_{1,\mathrm{N}}$, with contour length $L$ = 1845 Å and persistence length $p$=3.5 Å. The value of $L$ is slightly larger than the full length of the protein and $p$ is comparable to the bond length.

Experimental results for titin show serial unfolding, and are often fit to a sequence of WLC curves between each peak. The persistence lenth is constant, but as each domain unfolds the contour length $L$ increases by a fixed amount $\Delta L$. The dashed curves through the $\widetilde{T} = 0.3$ results for 5tit in Fig. 16 are fits of our calculated force curve to this common model. The fits use an initial length of $L$ = 620Å, $\Delta L$= 300Å, and $p$ = 3.5Å. Comparable fits can be obtained with a wide range of $p$ by shifting $L$ and using nearly the same $\Delta L$. For comparison, Rief et al.[13] show a fit with $\Delta L$ varying from 280 to 290Å and $p$= 4Å. The small value of $p$ indicates that titin is a nearly freely-jointed chain. After each peak the force drops to the new WLC curve. The value after the drop grows with the number of unfolded domains. The reason is that the new worm-like chain starts at a larger fraction of its fully extended length, and thus a larger force.

**Rate and γ dependence**

The results shown above were all for a fixed $v_p$=0.005Å/τ, but are only weakly dependent on $v_p$. In general, decreasing $v_p$ gives more time for thermal fluctuations and thus lowers the force peaks. This effect can be illustrated by changing the damping parameter γ, because decreasing γ is equivalent to lowering the effective pulling velocity (1/τ ~ γ in the overdamped limit). Figure 17 compares $F\!-\!d$ curves at $T$=0.2ε/$k_B$ for γ=2, 4, and 8 $m$/τ and the soft cantilever. While the velocity scales linearly with γ, the changes in the first force peak are small. As expected, the peak decreases and shifts to smaller $d$ as γ and the effective velocity decrease. The only place where the change in force is large is near the second force peak. This peak shifts rapidly with temperature. At $\widetilde{T} = 0.2$ there is just enough time for thermal activation at γ = 2, but not at the other γ. We have checked that further decreases of $v_p$ at γ = 2 produce little additional change in the second peak. Increasing $v_p$ leads to behavior that is closer to that at larger γ and $\widetilde{T} = 0$.

Experiments show a roughly logarithmic dependence of peak height on pulling rate.[13,24,40,47] This logarithmic dependence can be obtained from a wide range of models,[20,50] making it hard to deduce information about the energy landscape. Our results have implications for some of the models. For example, Makarov et al.[49] have provided a thorough discussion of rate dependence in a simple model of titin. They pointed out that the peak force should drop with decreasing rate because of thermal activation, and with increasing number of domains because of the increased sampling of rare events. Both effects are roughly logarithmic in their model. These key conclusions are likely to apply to any set of interactions, but their detailed calculations used a model that may be over-simplified. They assumed that only six A'+G bonds were important in the first peak, while Figure 15 implies that other bonds play a role, at least indirectly. A second assumption was that the six bonds fluctuated independently. We find that the elastic coupling along A' and G is so strong that thermal fluctuations do not break the main A'+G bonds independently. Indeed applying stress along the A' and G segments makes them tauter and less likely to fluctuate independently. Thus modeling their motion by a simple two-state model[20,24,40,47] may be more appropriate.

Zinober et al.[20] have made detailed comparisons between two-state models and experiments on extremely pure pentamers. Their results emphasize the importance of domain number in altering the compliance as well as increasing the number of potential failure points. The compliance changes with the number of unfolded domains, influencing the mechanical resistance of the next unfolding event. As they emphasize, the interconnected dependence of force peaks on the number of domains, cantilever stiffness and rate complicates the comparison between experiments by different groups and with simulations.

## SUMMARY AND CONCLUSIONS

The results presented here show that thermal fluctuations produce a variety of changes in the force needed to unfold proteins and the order of bond breaking. Some of these features are universal. For example, thermal fluctuations lead to activated rupture of bonds before they become mechanically unstable and thus lower the peaks in force–displacement curves. At high temperatures bond energies become irrelevant and all proteins act like purely entropic worm-like-chains.[42] The rate at which temperature produces these universal changes depends strongly on the geometry of the protein. Different geometries also lead to very different changes in the unfolding scenarios.

In $\alpha-$helices bonds between each pair of adjacent coils must transmit the pulling force and can unfold independently. The presence of many parallel points of failure leads to strong temperature dependence. At low temperatures, the bonds break in order of increasing strength. At higher temperatures, the difference in bond energies becomes unimportant. A fraction of bonds is broken at any pulling force, but each bond can fluctuate between broken and unbroken states. There is an interesting plateau in the force–displacement curve where broken and unbroken states are equally favored. This leads to a sort of coexistence between the two states which allows the protein to expand at fixed force much like a system with coexisting liquid and gas phases expands at fixed pressure.

In $\beta-$sheets, the entire force is focused on the unbroken bond closest to the external force, and bonds ultimately break in this order. Thermal fluctuations produce more gradual changes than for the $\alpha-$helix because there is only one potential site for failure. The nature of the unfolding scenario at intermediate temperatures depends on the stiffness of the pulling spring. When the spring is stiff, fluctuations in protein length are suppressed. At any given displacement, the terminal bond fluctuates between broken and unbroken states, but the protein unzips in a steady manner. The force curve shows a regular series of peaks and dips as the length varies. When the spring is soft, the protein length can vary without changing the force. As for the $\alpha-$helix, there is a force where broken and unbroken states are in balance and the length of the protein is nearly arbitrary. This leads to a long plateau in the force curve. The final breaking point of bonds has the same order, but all bonds break and reform many times.

The structure of titin is much more complex than the above examples. The simplicity and flexibility of the Go model, allowed us to examine the role of different sets of bonds in determining the maximum unfolding force. As in experiment and all-atom simulations,[24] removing bonds to the A strand had little effect on the peak height. Most of the stress was carried through A'+G bonds as concluded from all-atom simulations[22,23]. While other bonds had little direct effect on stress transmission, they did affect the peak force by stabilizing the geometry of the A' and G strands.

As in the β sheet, most of the bonds in titin are shielded from the pulling force. However, in contrast to the simple secondary structures, the A'+G bonds in titin are initially perpendicular to the force and can all carry the load in parallel. This leads to much larger peak forces, and suppresses the effect of thermal fluctuations. As shown in Fig. 11, peaks are still observed in the force–displacement curves at $\tilde{T} = 0.6$ which is just below $\tilde{T}_{max}$.

Force curves for multiple domains of titin reproduce many features of experimental results. At low temperature the unfolding is predominantly serial. However, there is a shoulder before the first peak where each domain transforms from the native state to an intermediate state by breaking A+G bonds. As in Ref. 24, the length of this shoulder grows linearly with the number of domains. At intermediate temperatures, the bulk of the force curve can be fit to a sequence of WLC curves following the approach used to fit experimental data[13] and with similar contour and persistence lengths. At high temperatures the entire force curve follows that of a single, long WLC.

The results presented here were mainly for a single velocity. Recent experiments suggest that formation of the intermediate state may be suppressed at velocities below 1nm/s.[40] Such velocities are not directly accessible to simulations, but it may be possible to observe this change in behavior by varying $\tilde{T}$. The effect of velocity on the second peak in titin would also be an interesting subject for future studies.

## FIGURE CAPTIONS

**Fig. 1.** Initial stages of stretching for a single domain of titin at the indicated $\tilde{T}$. The solid lines correspond to averaging over 0.5 Å or 100 τ and the dotted lines to 0.05 Å or 10 τ.

**Fig. 2.** Force–displacement curves for three Go models at $\tilde{T} = 0$. The solid lines are for the model considered in this paper. It uses the contact map determined from atomic van der Waals radii and 6-12 Lennard-Jones contact potentials. The dotted line in the bottom panel shows the effect of changing from a 6-12 to 10-12 Lennard-Jones contact potential. The dotted line in the top panel shows results for the contact map and tethering potential used previously[21,27].

**Fig. 3.** Force-displacement curves for the Go model of the α-helix H16 with the stiff springs attached to it. The dashed line in the top panel corresponds to $\tilde{T} = 0$, and the other two lines show two different trajectories at $\tilde{T} = 0.1$. In the bottom panel, the solid, dotted, and dashed lines correspond to $\tilde{T} = 0.2$, 0.4, and 0.6, respectively. The force is averaged over a distance of $d_A = 0.5$ Å.

**Fig. 4.** Same as in Figure 3 but for the soft spring case.

**Fig. 5.** Plots of the maximum force (top panel) and the corresponding tip displacement (bottom panel) for the stiff (circles) and soft (squares) springs.

**Fig. 6.** Force-displacement curves for the Go model of the β-hairpin B16. The top and bottom panels are for the stiff spring case at the indicated $\tilde{T}$. The inset shows the dependence of the maximum force on $\tilde{T}$ for stiff (squares) and soft (circles) springs.

**Fig. 7.** Force-displacement curves for the Go model of the β-hairpin B16 with soft springs at the indicated $\tilde{T}$.

**Fig. 8.** The ribbon representation of the I27 domain, coded as 1tit in the Protein Data Bank. The symbols indicate particular β-strands, together with the sequence position of the amino acids in each.

**Fig. 9.** The main figure shows the median folding time for 1tit, and the arrow indicates the folding temperature $\tilde{T}_f$. The inset shows the specific heat in units of $k_B$ as a function of $\tilde{T}$.

**Fig. 10.** The succession of folding events in the Go model of 1tit as illustrated by times needed to establish a contact versus the contact order. The letter symbols indicate the nature of the event, e.g. the open circles indicate formation of contacts between strands B and G. The asterisks show contacts which do not correspond to two strands (but, say, to an unstructured fragment and a strand).

**Fig. 11.** The temperature dependence of *F–d* patterns in 1tit for the soft spring case. The association of the data lines with the temperatures is as follows: the solid thick line – $\tilde{T}=0$, the dotted line – $\tilde{T} = 0.2$, the thin solid line – $\tilde{T} = 0.6$.

**Fig. 12.** Similar to Figure 11 but for the stiff pulling spring. Data for $\tilde{T} = 0.6$ are not shown for clarity.

**Fig. 13.** Heights of the two main force peaks (top) and their positions (bottom) as a function of $\tilde{T}$. Open and closed symbols refer to the stiff and soft pulling springs, respectively.

**Fig. 14.** Contact breaking distances versus the contact order for the soft (top) and stiff (bottom) springs at $\tilde{T} = 0$, 0.2 and 0.8. The left panels define symbols that are used to indicate bonds connecting each pair of the native strands defined in Fig. 8. Contacts where at least one of the amino acids is not part of a native strand are indicated by asterisks.

**Fig. 15.** The variation of the force near the first peak for titin at $\tilde{T} = 0$ with stiff springs and different sets of native contacts included. The dash-dotted line shows the force with all native contacts included. The dotted line shows results when native contacts involving the A strand are eliminated. The solid line shows results for A'+G, A'+A' and G+G bonds only. For the dashed line, all bonds involving only A', B, F and G are included.

**Fig. 16.** Force vs. end-to-end length of protein $d_{1,\,N}$ for a quintuple tandem repeat of the I27 domain of titin at the indicated temperatures with soft springs. Smooth dashed lines show WLC fits for $\widetilde{T} = 0.3$ and 0.8. Data for $\widetilde{T} = 0.8$ were averaged over 10 consecutive points so that the fit can be seen.

**Fig. 17.** The dependence of $F$–$d$ curves on the damping parameter $\gamma$ with soft springs. The dotted line is for the value $\gamma = 2m/\tau$ used elsewhere in this paper. The thin and thick solid lines are for $\gamma = 4$ and 8 $m/\tau$, respectively.

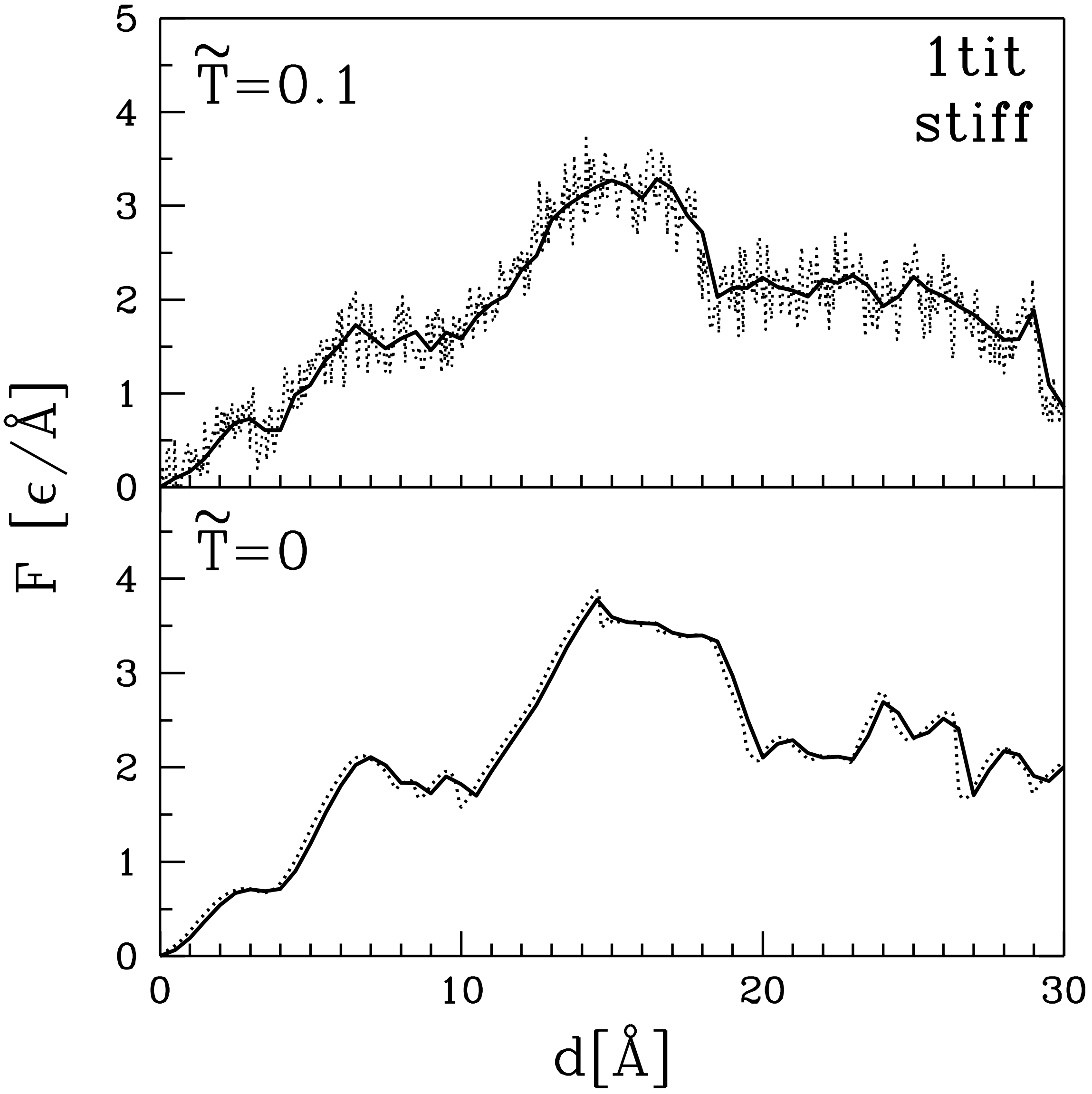

Figure 1.

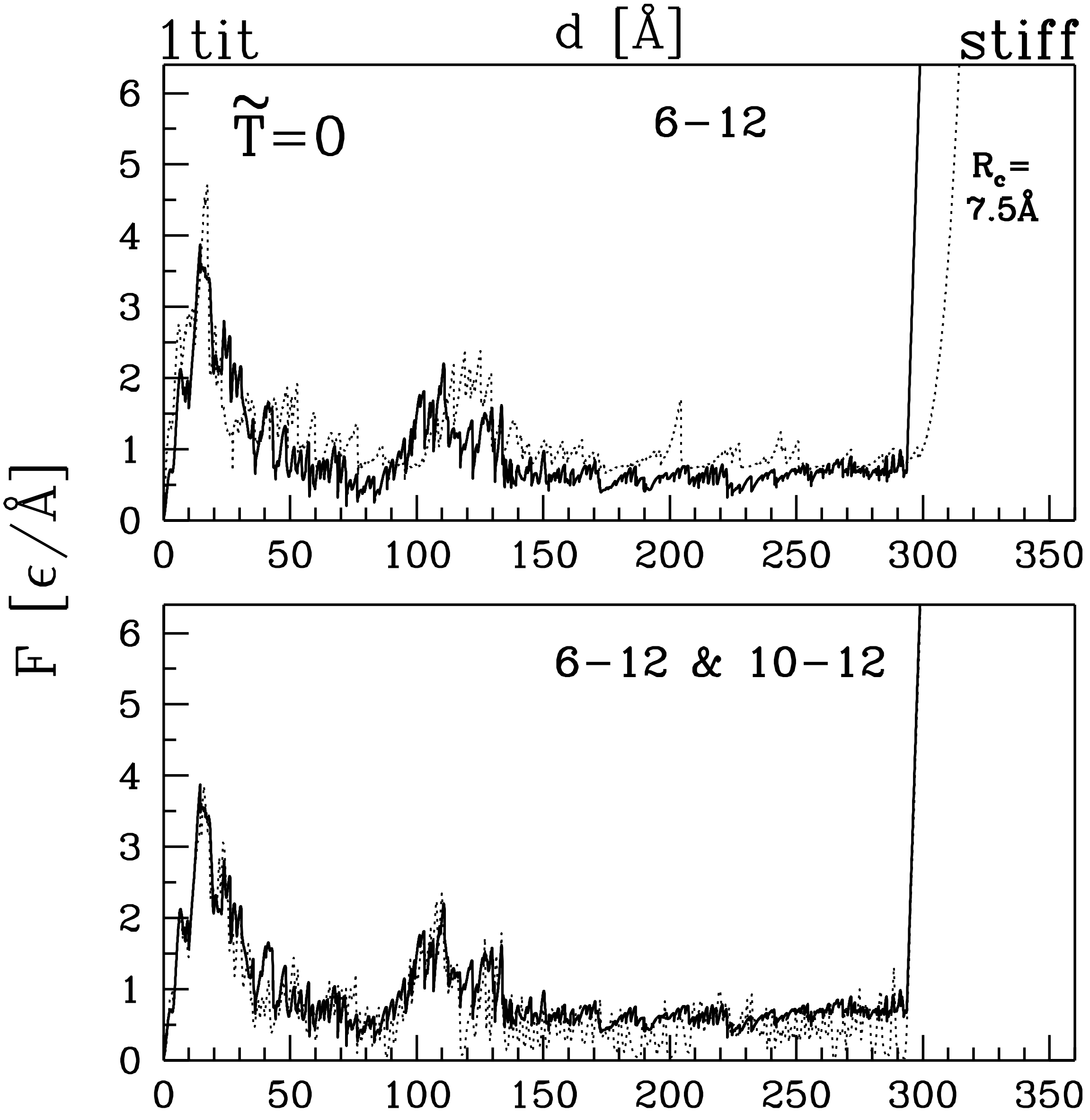

Figure 2

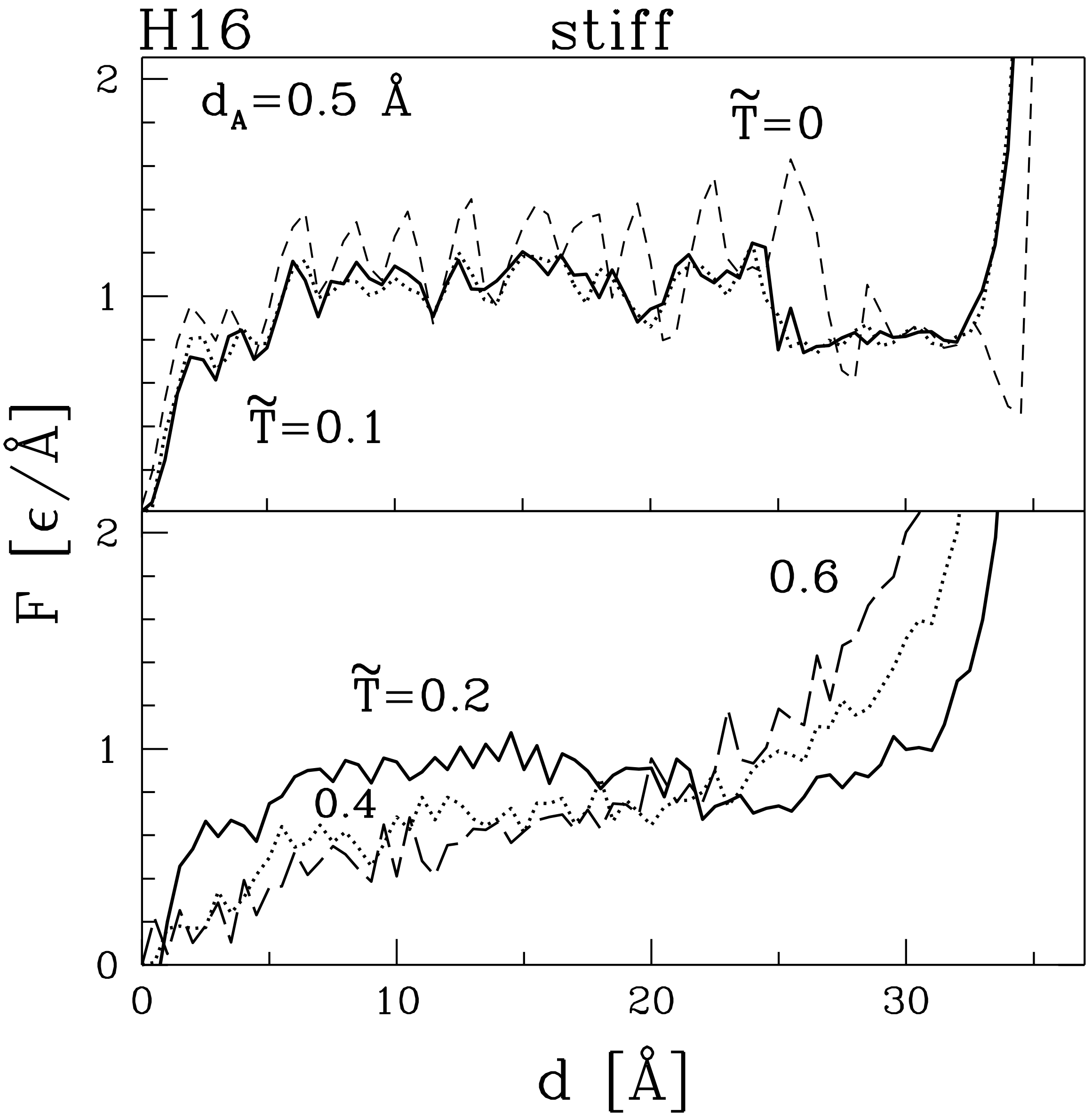

Figure 3

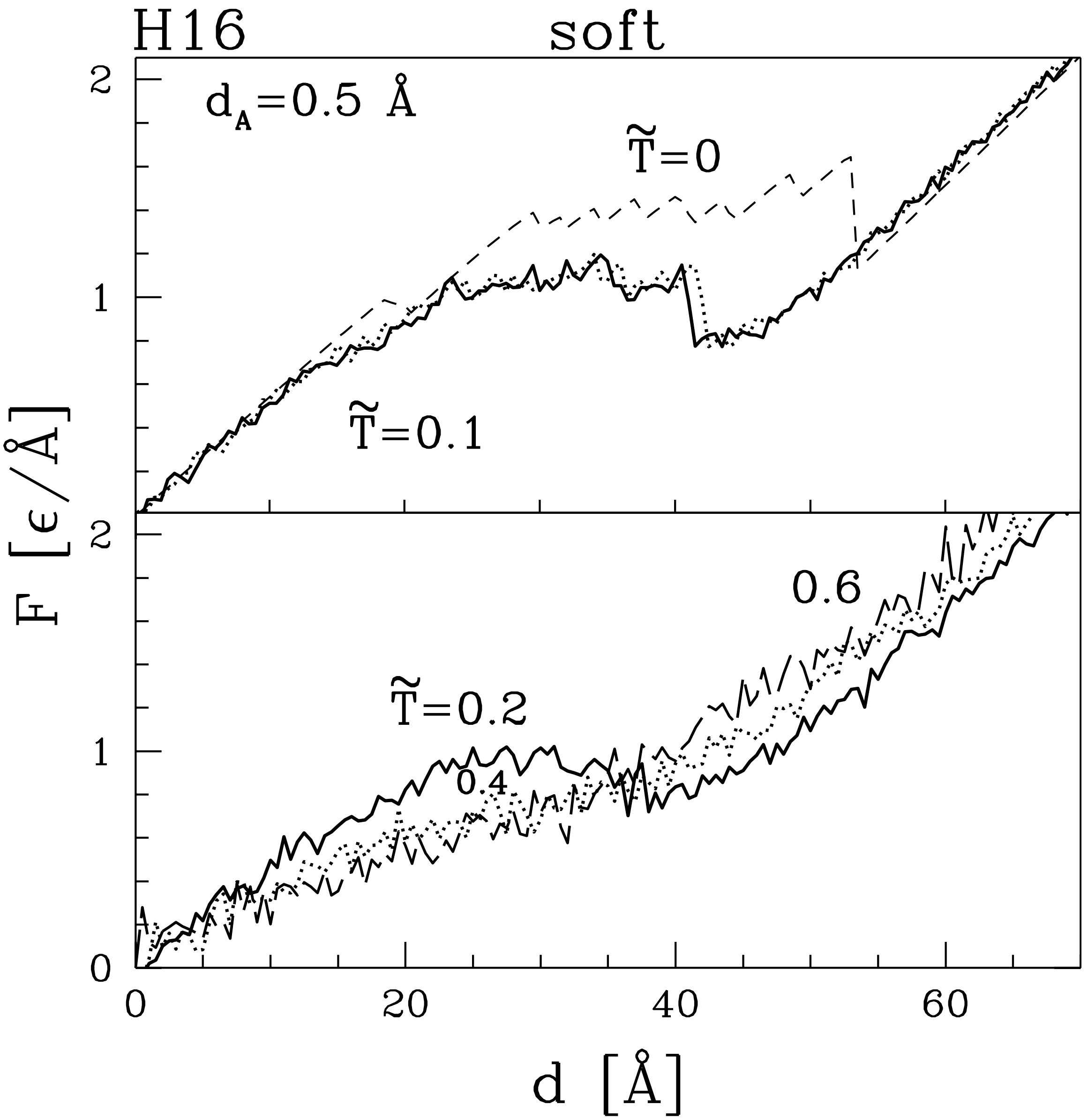

Figure 4

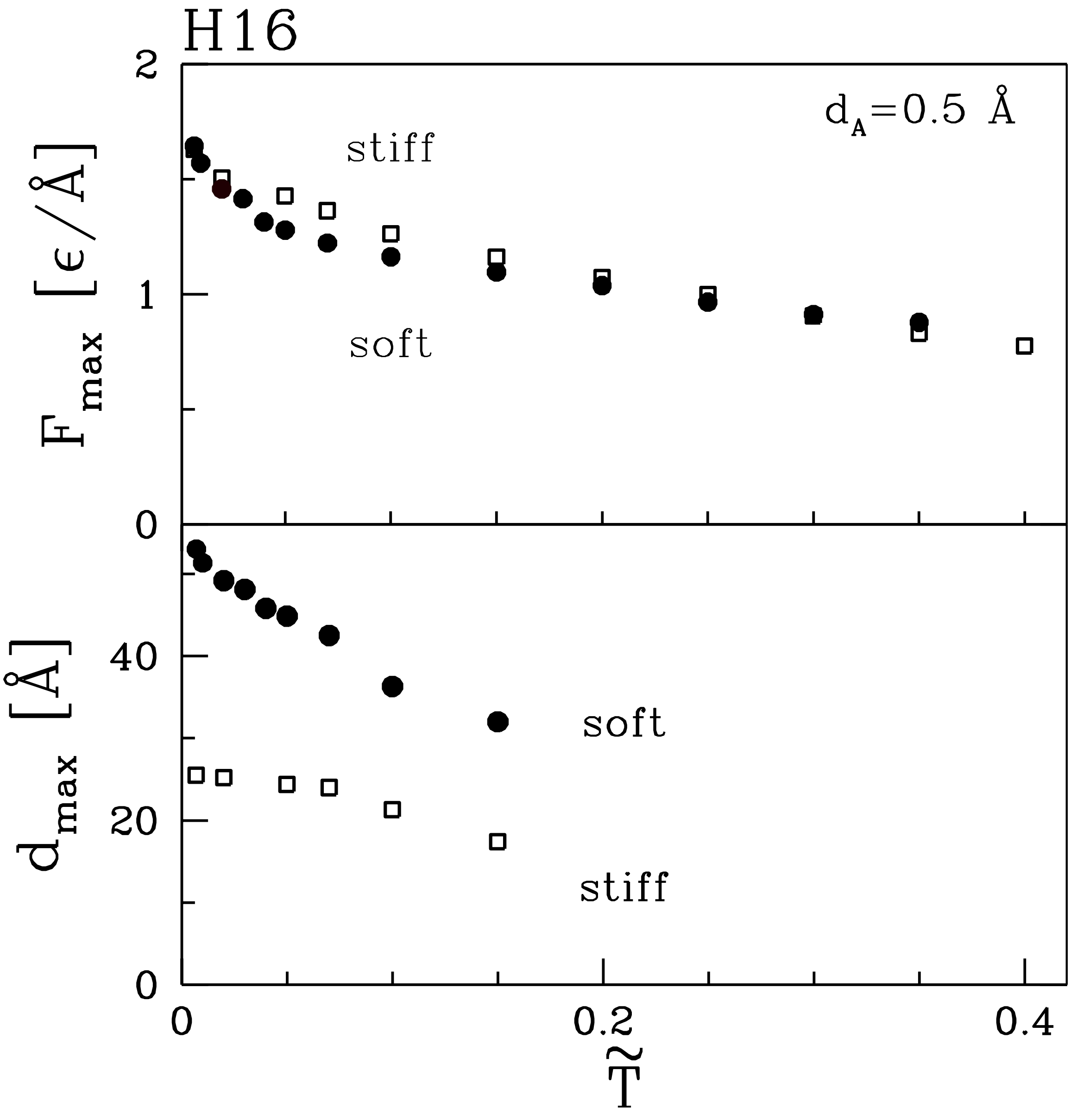

Figure 5

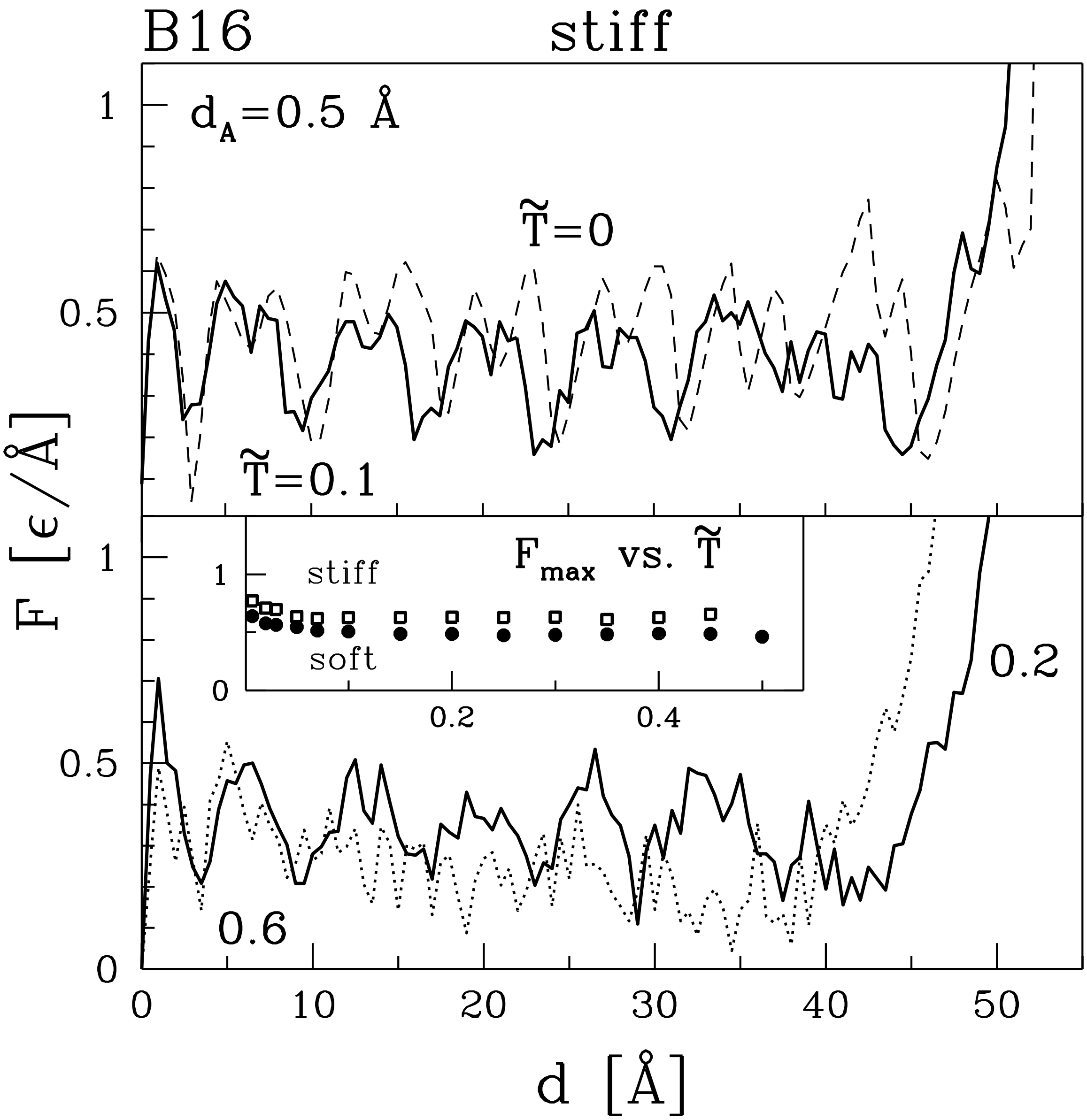

Figure 6

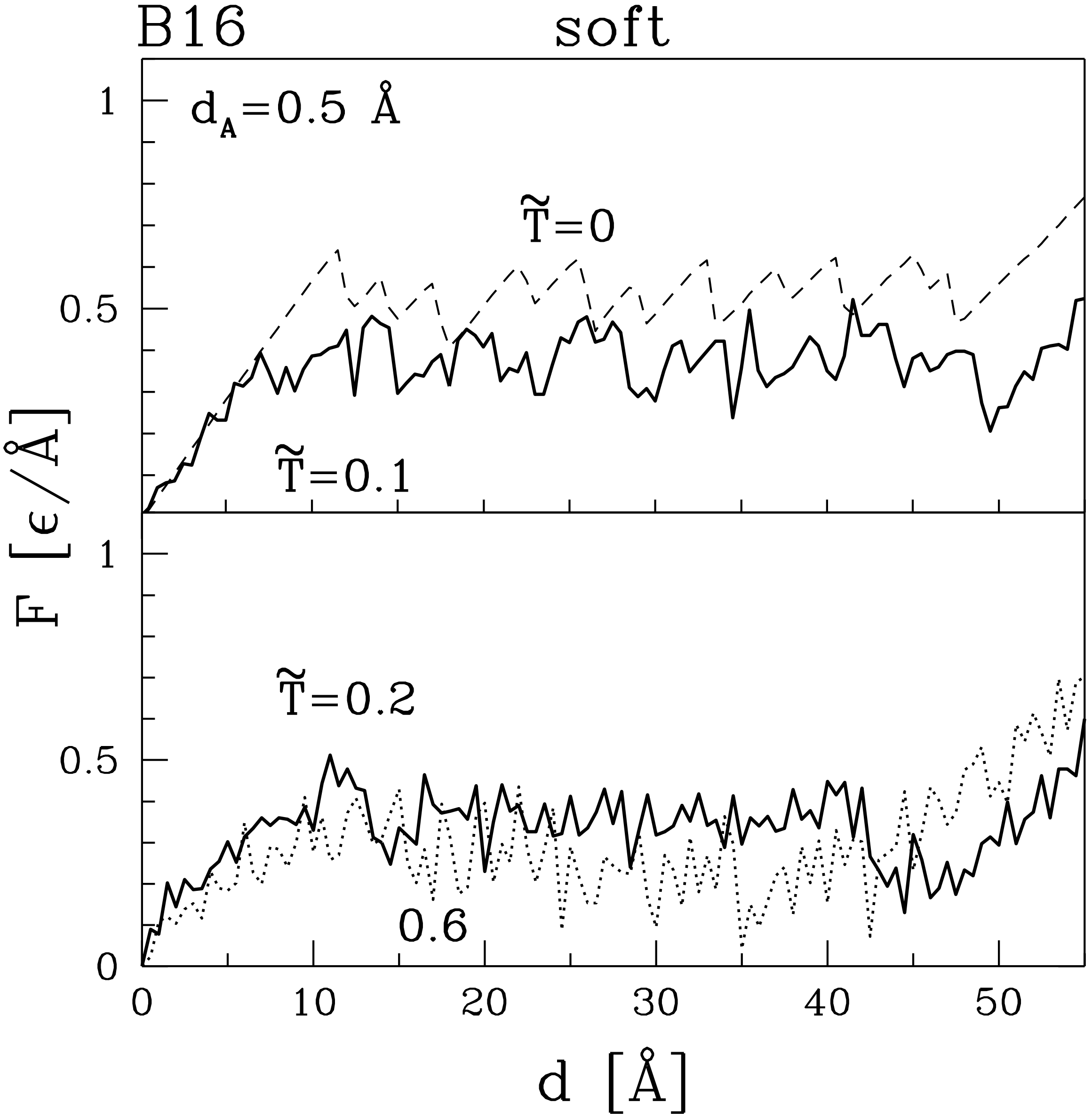

Figure 7

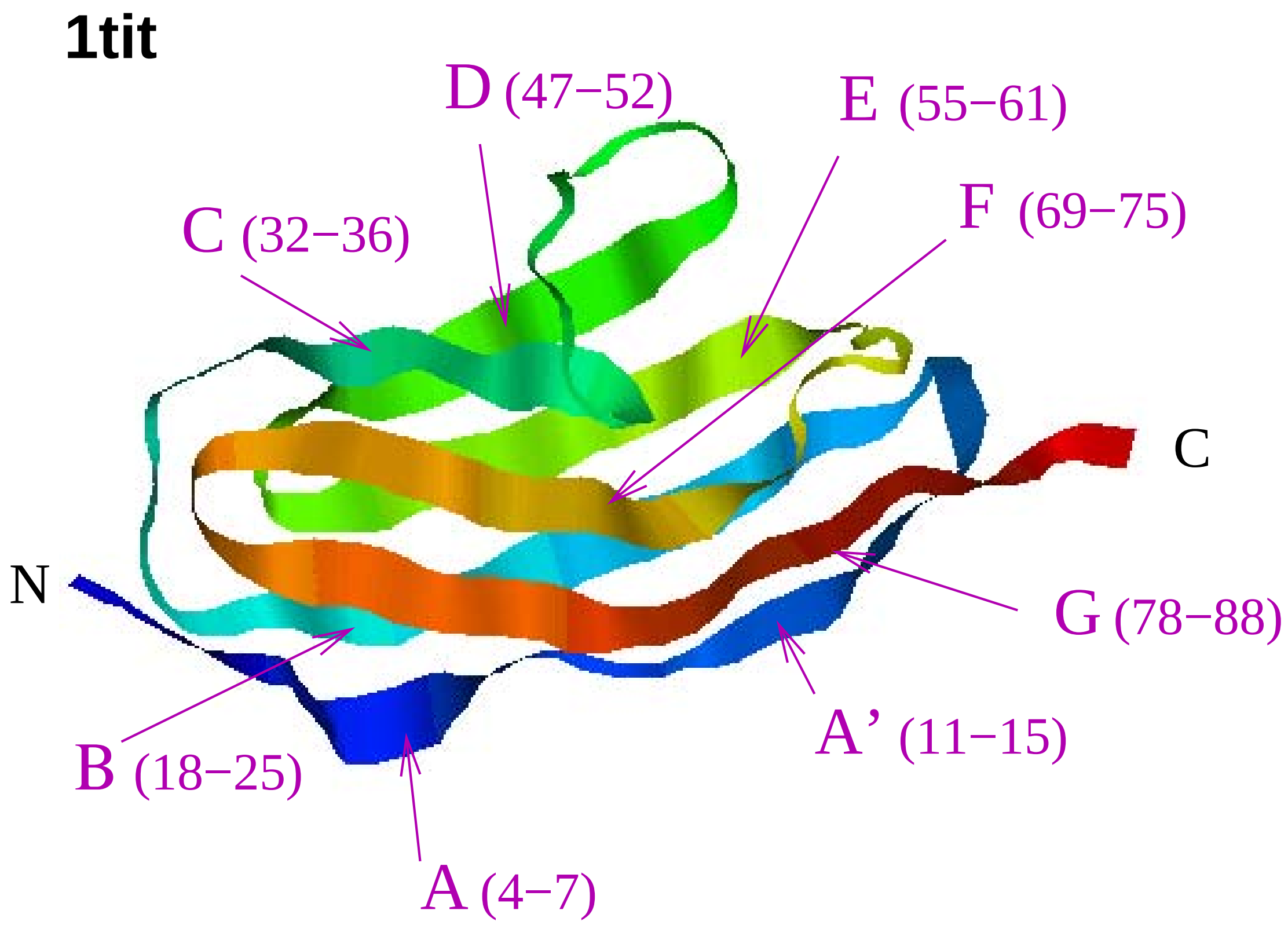

Figure 8

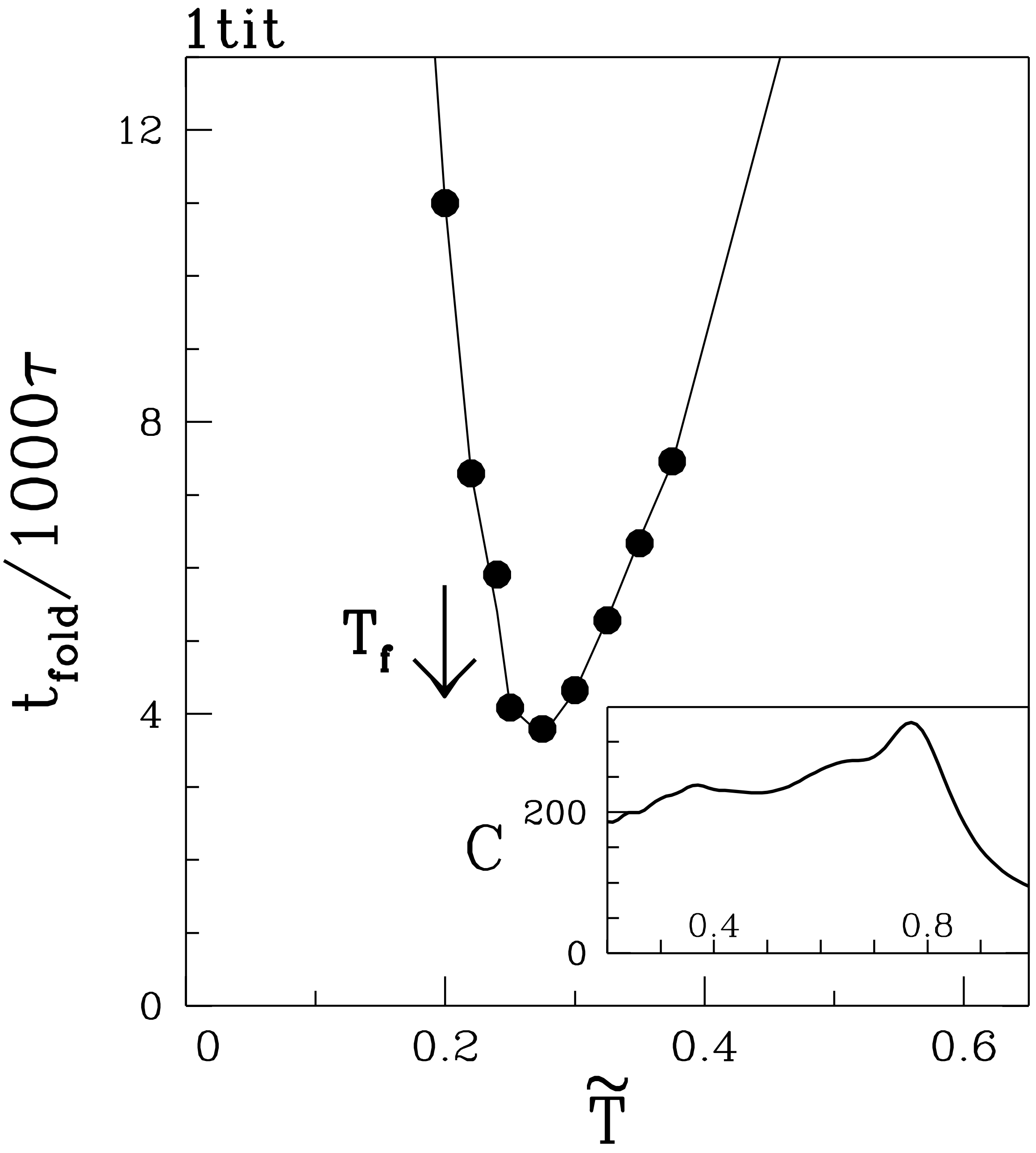

Figure 9

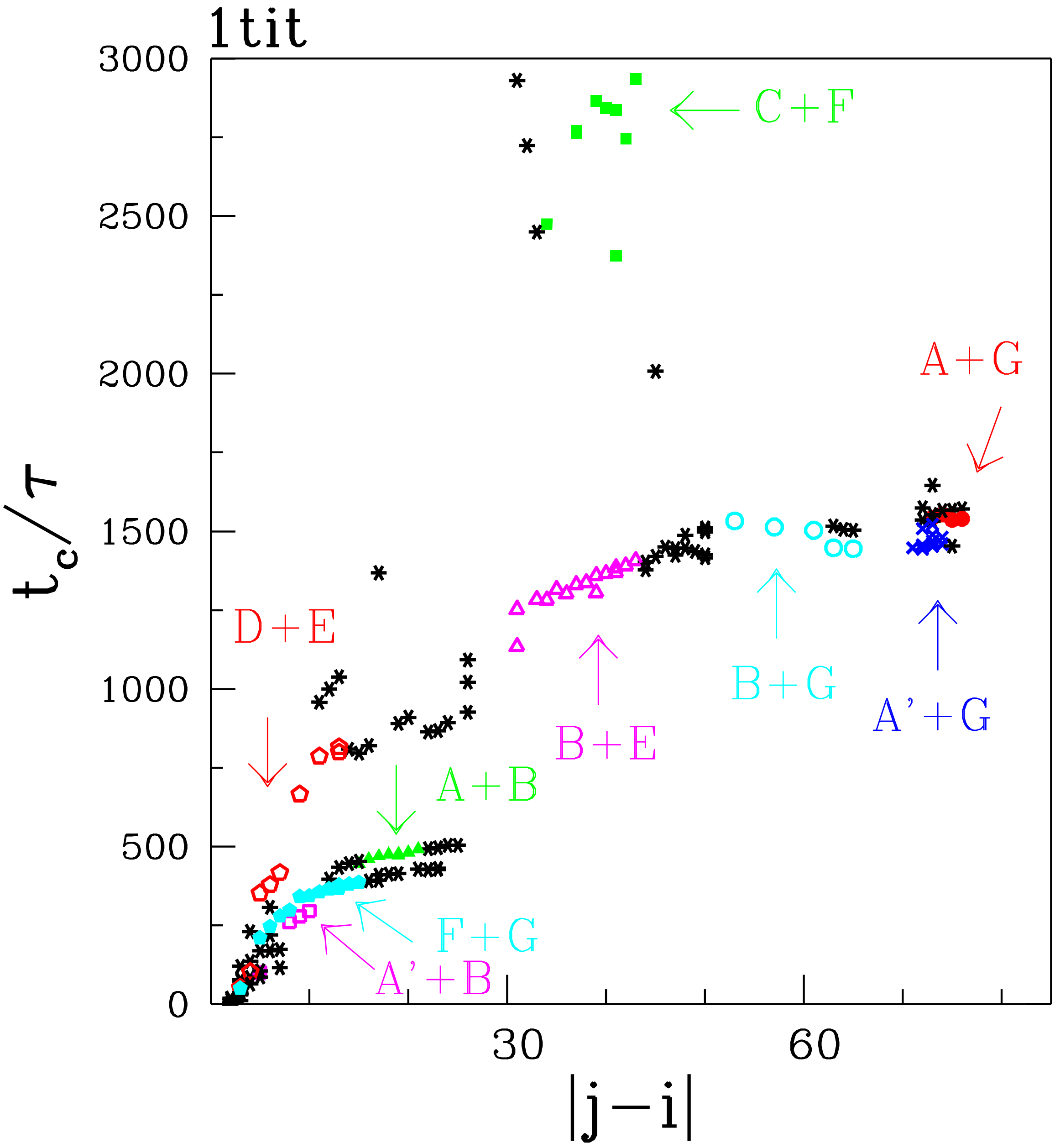

1tit
3000
2500
2000
1500
1000
500
0
30
60
$t_c/\tau$
|j−i|
C+F
A+G
D+E
B+E
B+G
A'+G
A+B
F+G
A'+B

Figure 10

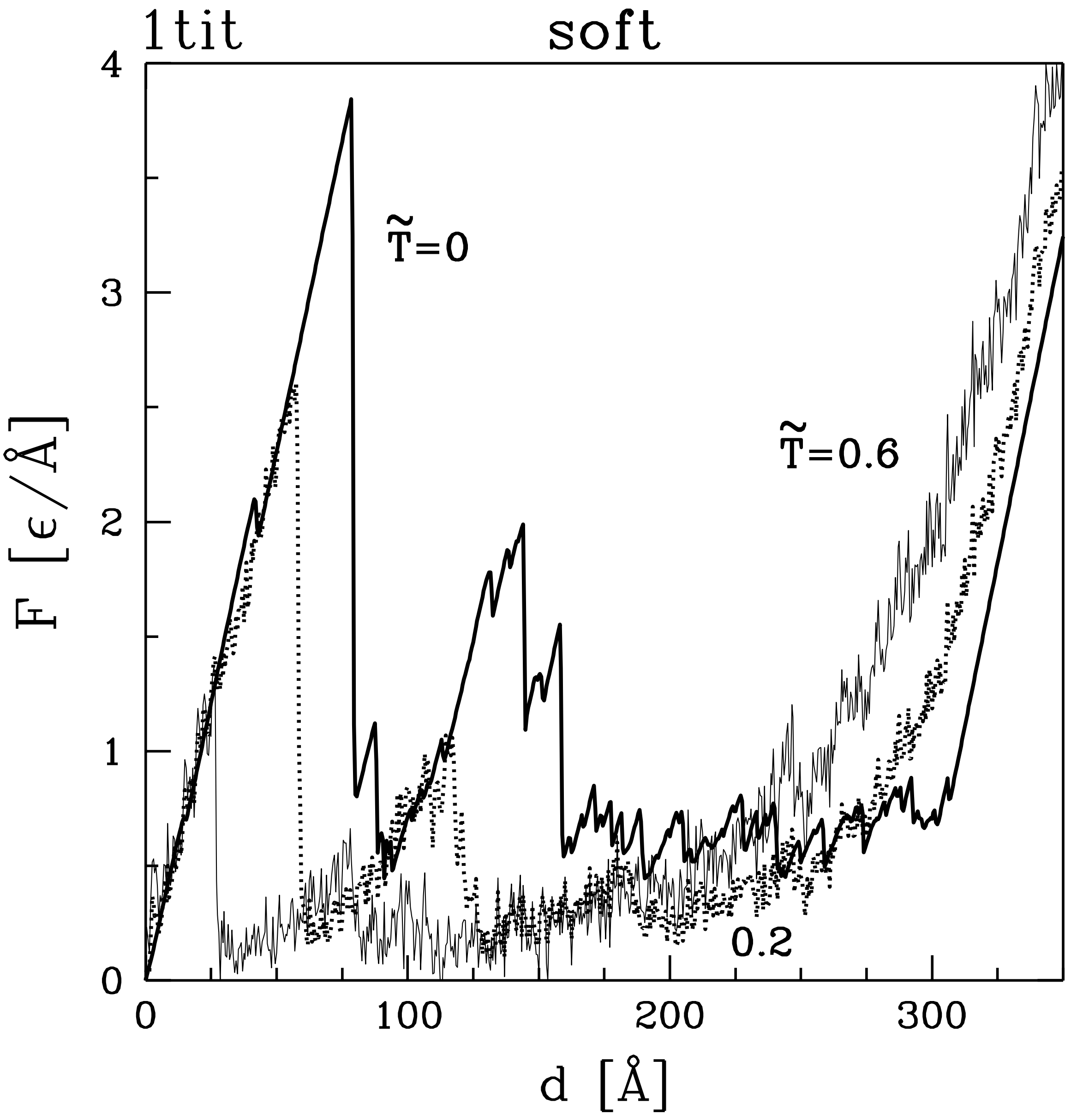

Figure 11

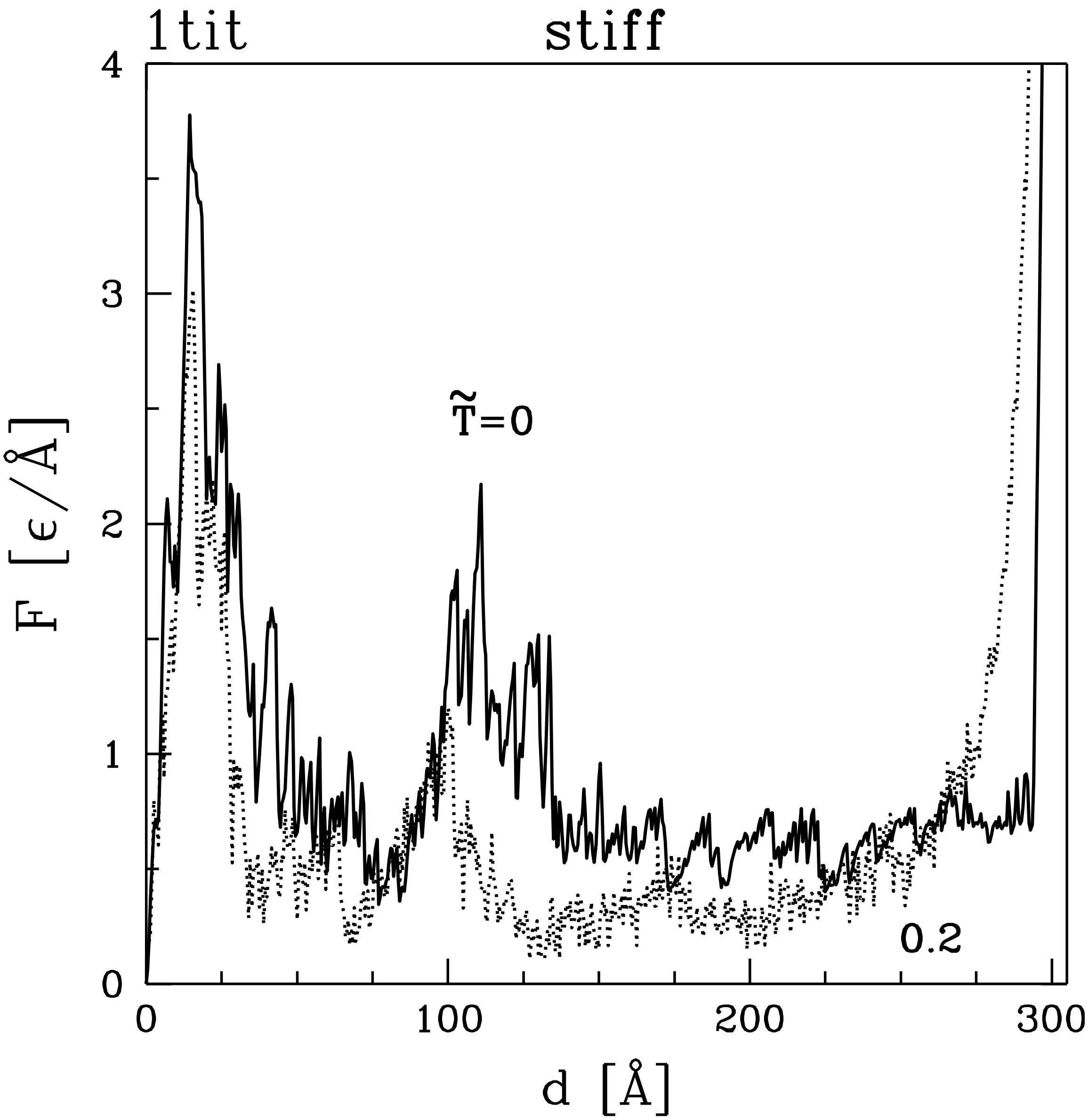

Figure 12

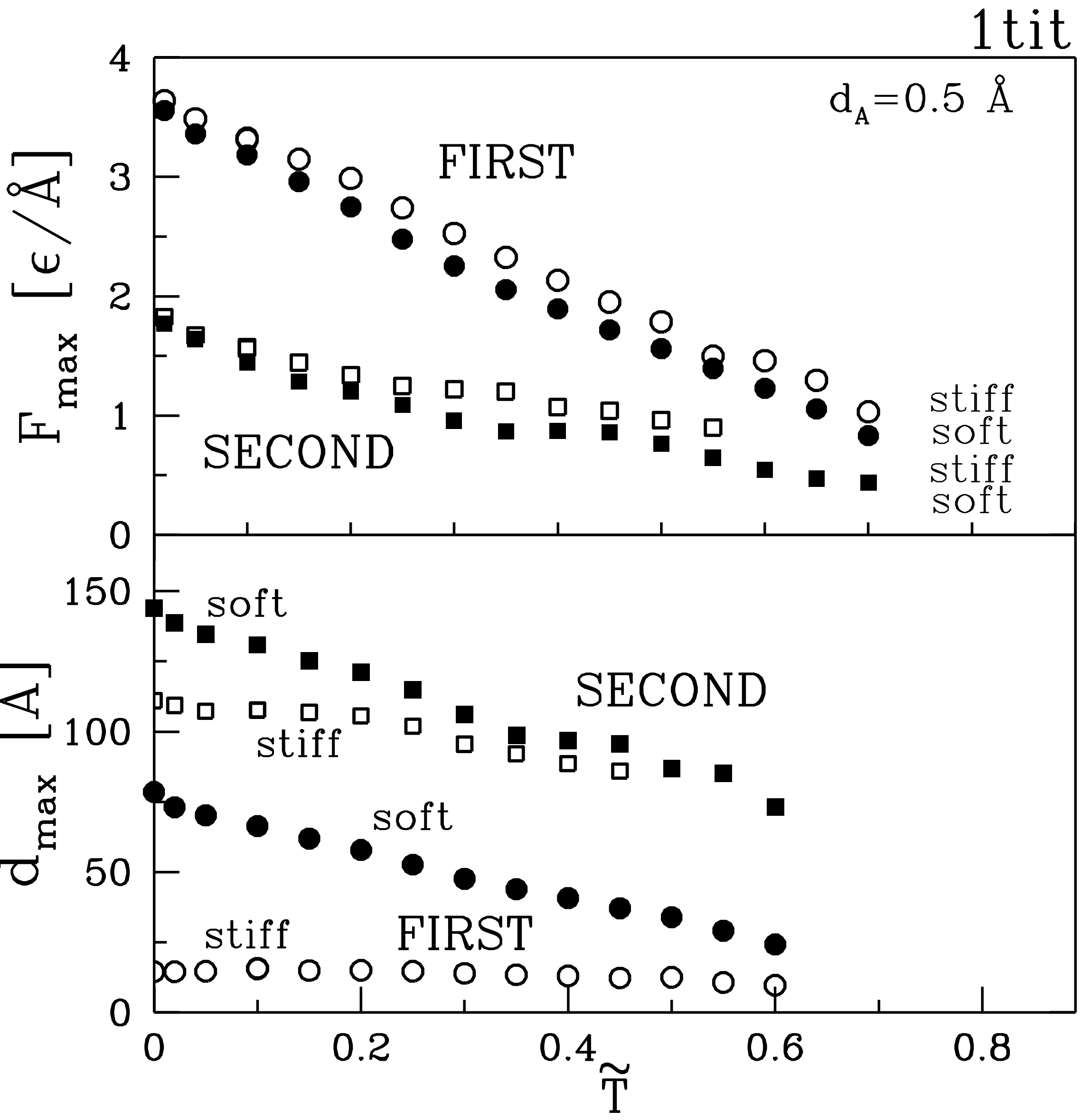

Figure 13

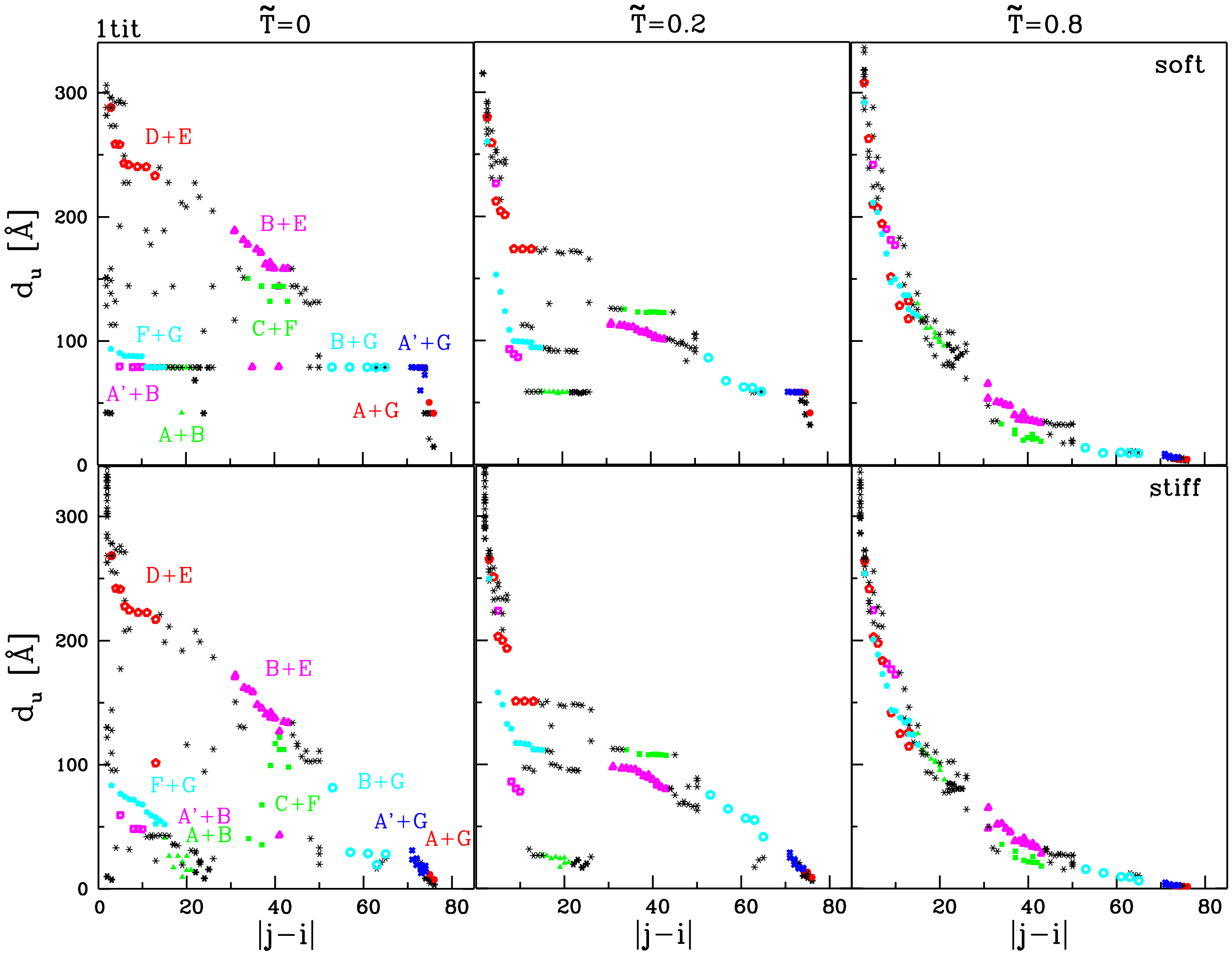
1tit
$\tilde{T}$=0
$\tilde{T}$=0.2
$\tilde{T}$=0.8
soft
stiff
$d_u$ [Å]
|j−i|
D+E
B+E
C+F
F+G
B+G
A'+G
A'+B
A+B
A+G

Figure 14

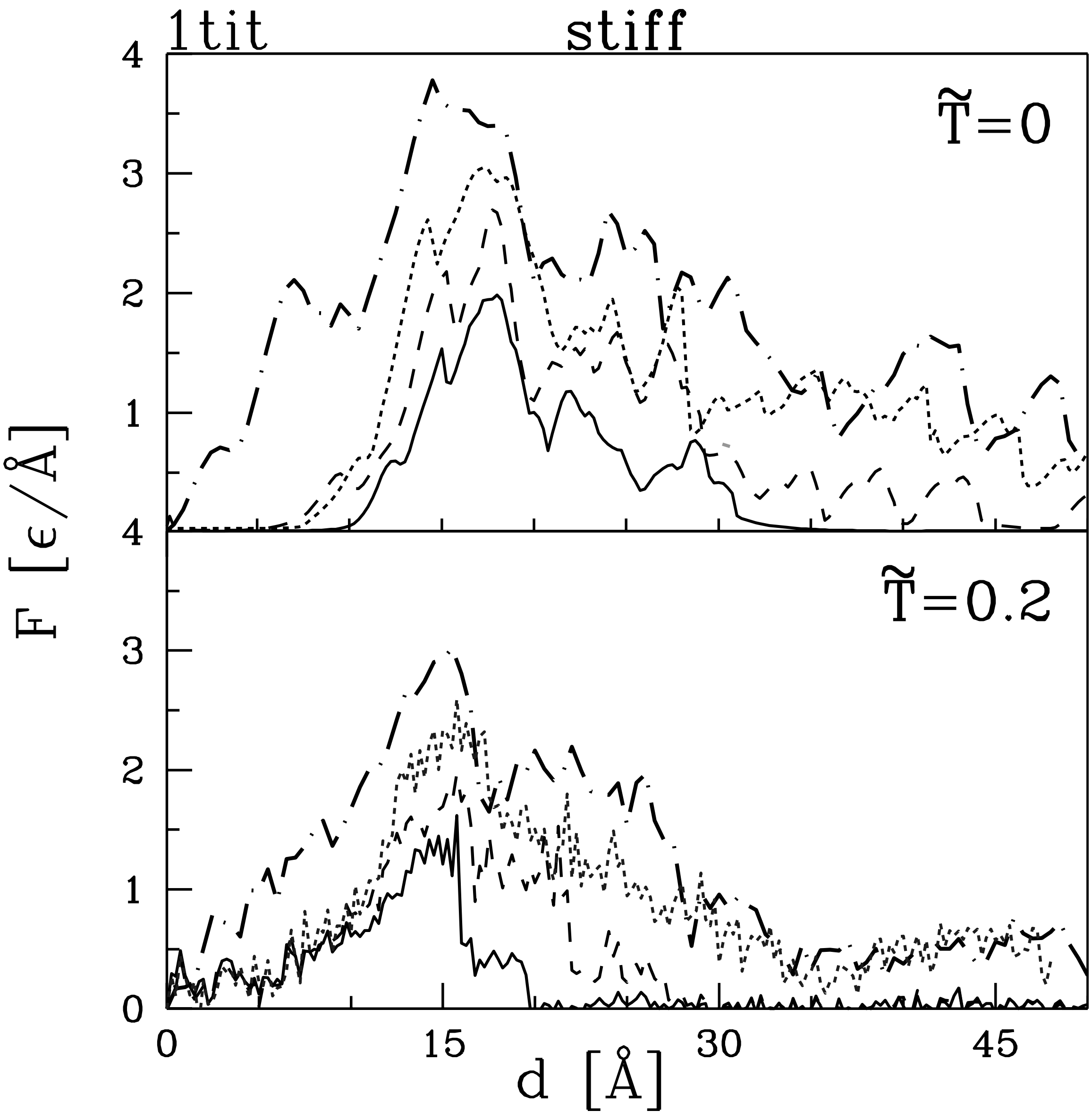

Figure 15

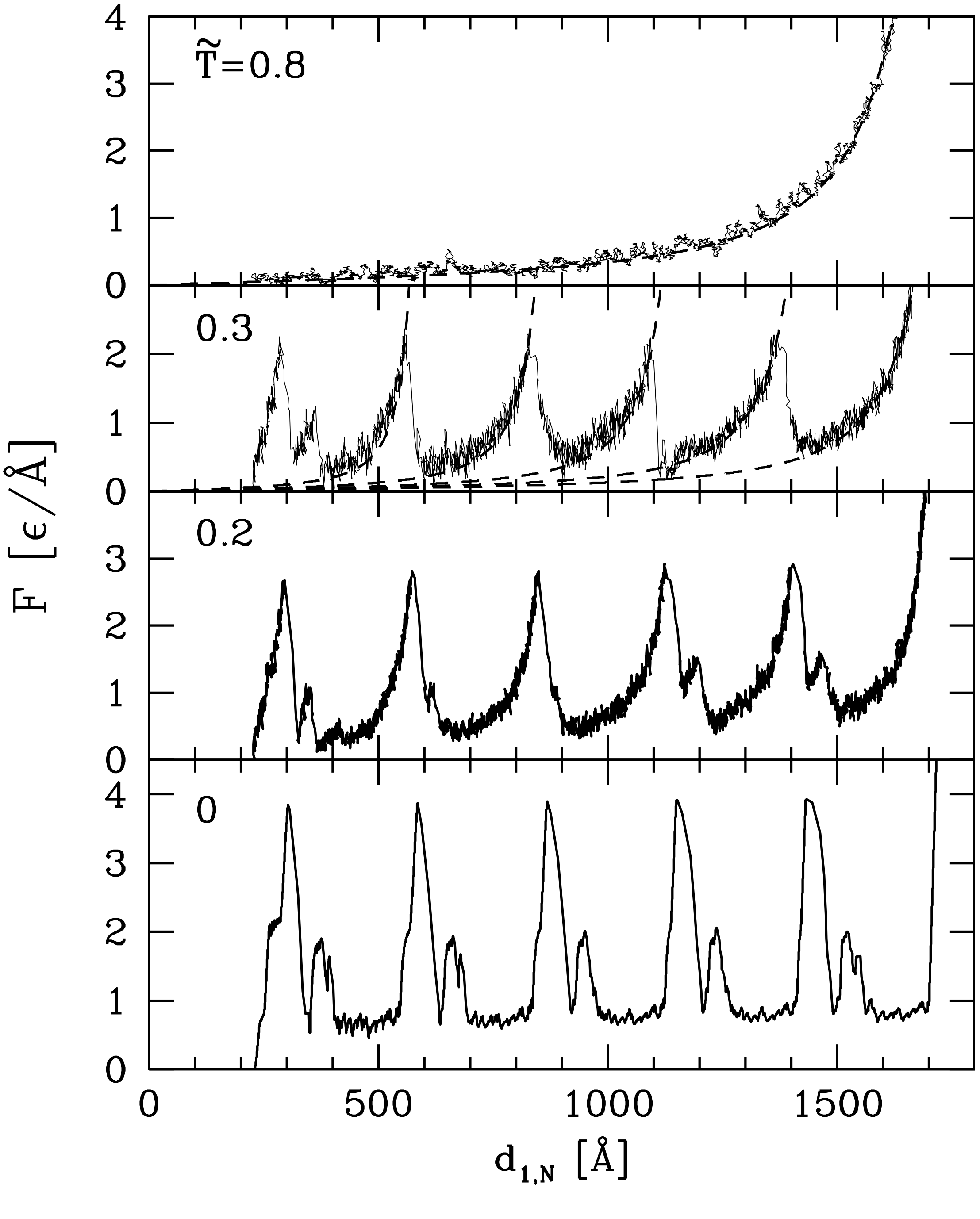

Figure 16

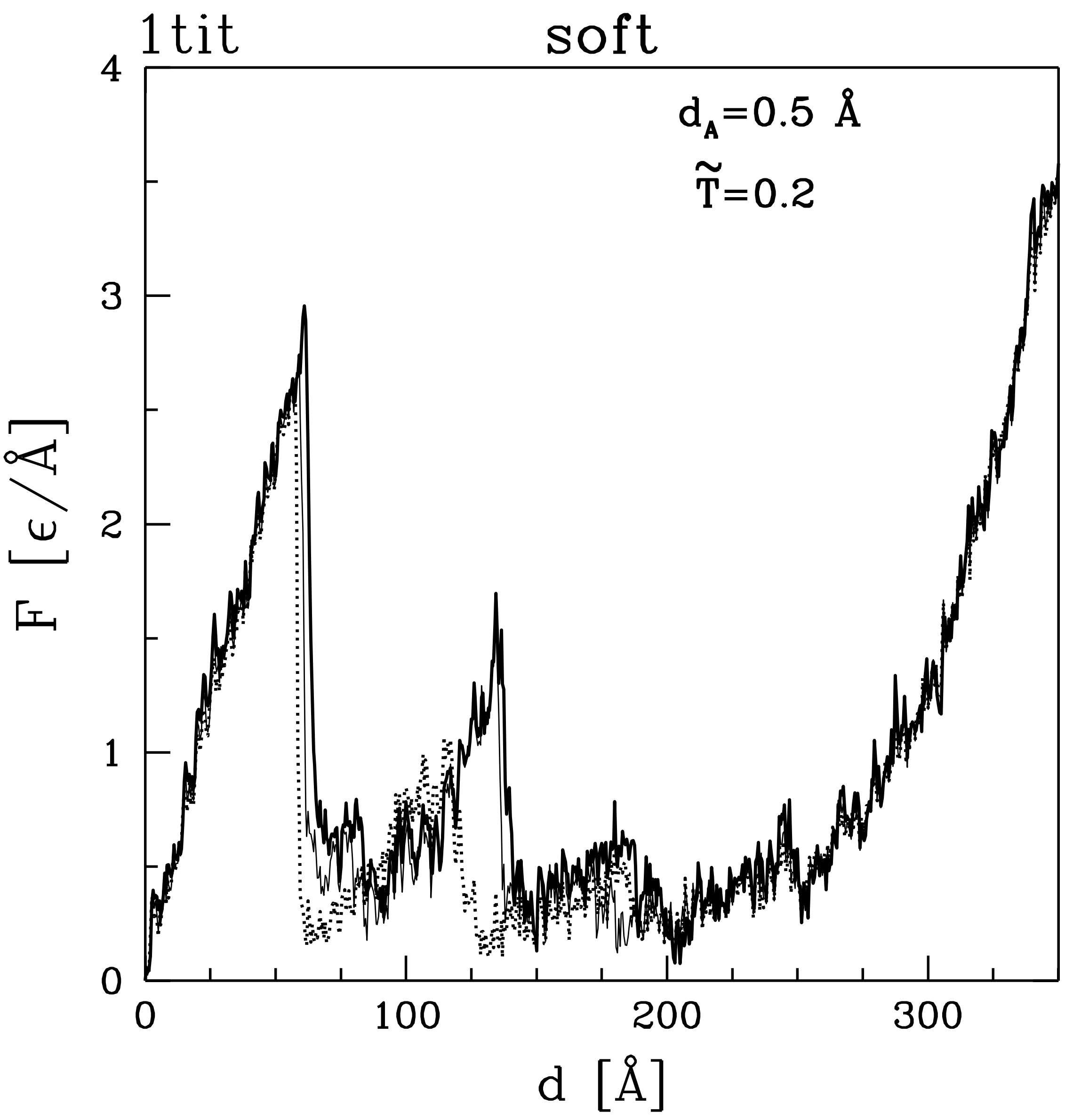

Figure 17